\documentclass[a4paper,fleqn]{article}
\usepackage{amsmath}

\begin{document}

\title{\textbf{Integrability of one bilinear equation: singularity analysis and dimension}}

\author{\textsc{Sergei Sakovich}\bigskip \\
\small Institute of Physics, National Academy of Sciences of Belarus \\
\small sergsako@gmail.com}

\date{}

\maketitle

\begin{abstract}
The integrability of a four-dimensional sixth-order bilinear equation associated with the exceptional affine Lie algebra $D_4^{(1)}$ is studied by means of the singularity analysis. This equation is shown to pass the Painlev\'{e} test in three distinct cases of its coefficients, exactly when the equation is effectively a three-dimensional one, equivalent to the BKP equation.
\end{abstract}

\section{Introduction}

Most of the integrable nonlinear partial differential equations of modern mathematical physics are two-dimensional ones. There are very few three-dimensional integrable nonlinear equations in the literature. And even less is known about the integrability in dimension four and higher. In the present paper, we study one four-dimensional nonlinear equation and show that, unfortunately, this equation is integrable only if it is effectively three-dimensional.

We study the following four-dimensional sixth-order bilinear equation:
\begin{gather}
\left( D_x^6 + 36 D_x D_t - 10 D_y^2 - 10 D_z^2 \right) \tau \cdot \tau \notag \\
+ a \left( D_x^3 D_y + D_y^2 - D_z^2 \right) \tau \cdot \tau + b \left( D_x^3 D_z - 2 D_y D_z \right) \tau \cdot \tau = 0 , \label{e1}
\end{gather}
where $D_x$, $D_t$, $D_y$ and $D_z$ denote Hirota's bilinear differentiation operators, while $a$ and $b$ are parameters. This bilinear equation, associated with the exceptional affine Lie algebra $D_4^{(1)}$, appeared for the first time in \cite{KW89} (see p.~207 there). The same equation appeared recently in \cite{CVY18}, in a different form related to \eqref{e1} by a linear transformation of independent variables. It is interesting to investigate whether there are such values of the parameters $a$ and $b$, for which this bilinear equation \eqref{e1} is a genuine four-dimensional integrable equation.

In Sections \ref{s2}, we study the integrability of the four-dimensional sixth-order bilinear equation \eqref{e1} by the method of singularity analysis, in its formulation for partial differential equations \cite{WTC83,T89}. In Section \ref{s3}, we apply the same Painlev\'{e} test for integrability to other $D_4^{(1)}$-associated bilinear equations, initially three-dimensional ones. Section \ref{s4} contains discussion of the results.

\section{Singularity analysis} \label{s2}

In our experience, the Painlev\'{e} test for integrability is a reliable and easy-to-use method, especially convenient (if compared to other methods) for high-dimensional, high-order, non-evolutionary and multi-component nonlinear equations \cite{S94a,S94b,S95,S97,S98,KS01,KSY03,KS05,S05,S11,S13a,S17}. The reliability of the Painlev\'{e} test has been empirically verified by the integrability studies of fifth-order KdV-type equations \cite{HO85}, bilinear equations \cite{GRH90}, coupled KdV equations \cite{K97,S99a,S01b,S14}, coupled higher-order nonlinear Schr\"{o}dinger equations \cite{ST00a}, generalized Ito equations \cite{KKS01}, sixth-order nonlinear wave equations \cite{KKSST08}, seventh-order KdV-type equations \cite{X14}, etc. Some interesting new equations have been discovered by means of the singularity analysis \cite{KKS04,ST00b,ST00c,K08,S08,S19}. Also, the Painlev\'{e} property is very sensitive to the dimension and geometry of a studied nonlinear problem \cite{KS05,DS85} (see, however, a remark in \cite{S92} on why \cite{DS85} was criticized).

In usual partial derivatives and with the notation $\tau = u (x,y,z,t)$, the bilinear equation \eqref{e1} takes the form
\begin{gather}
 u u_{xxxxxx} - 6 u_x u_{xxxxx} + 15 u_{xx} u_{xxxx} - 10 u_{xxx}^2 \notag \\
 + 36 u u_{xt} - 36 u_x u_t - 10 u u_{yy} + 10 u_y^2 - 10 u u_{zz} + 10 u_z^2 \notag \\
 + a \left( u u_{xxxy} - u_y u_{xxx} - 3 u_x u_{xxy} + 3 u_{xx} u_{xy} + u u_{yy} - u_y^2 - u u_{zz} + u_z^2 \right) \notag \\
 + b \left( u u_{xxxz} - u_z u_{xxx} - 3 u_x u_{xxz} + 3 u_{xx} u_{xz} - 2 u u_{yz} + 2 u_y u_z \right) = 0 . \label{e2}
\end{gather}
A hypersurface $\phi (x,y,z,t) = 0$ is non-characteristic for this partial differential equation \eqref{e2} if $\phi_x \ne 0$, and we take $\phi_x = 1$,
\begin{equation}
\phi = x + \psi (y,z,t) , \label{e3}
\end{equation}
without loss of generality. Substitution of the expansion
\begin{equation}
u = u_0 (y,z,t) \phi^p + \dotsb + u_r (y,z,t) \phi^{p + r} + \dotsb \label{e4}
\end{equation}
to the nonlinear equation \eqref{e2} determines the admissible leading exponents $p$ (the dominant behavior of solutions $u$ near $\phi = 0$), as well as the corresponding resonances $r$ (the positions, where arbitrary functions can enter the expansion). In this way, we obtain the following two branches to be studied:
\begin{equation}
p = 1 , \qquad r = -1, 0, 1, 2, 3, 10 , \label{e5}
\end{equation}
and
\begin{equation}
p = 2 , \qquad r = -2, -1, 0, 1, 5, 12 , \label{e6}
\end{equation}
with $u_0 (y,z,t)$ being an arbitrary function in both cases. Let us note that the nonlinear equation \eqref{e2} does not possess branches with negative values of $p$ (this is typical for bilinear equations \cite{GRH90}), and that the expansions \eqref{e4} with \eqref{e5} or \eqref{e6} are not covered by the Cauchy--Kovalevskaya theorem because the Kovalevskaya form of the partial differential equation \eqref{e2} is singular at $u = 0$.

The branch \eqref{e5} is the generic one. In this case, the expansion \eqref{e4} represents the general solution of \eqref{e2} and (potentially) contains six arbitrary functions of three variables: $\psi (y,z,t)$, which corresponds to $r = -1$, and $u_i (y,z,t)$ with $i = 0, 1, 2, 3, 10$. The branch \eqref{e6} represents only a class of special solutions, because the expansion \eqref{e4} can contain only five arbitrary functions of three variables in this case: $u_0$, $u_1$, $u_5$, $u_{12}$ and $\psi$. For this reason, it seems natural to study the generic branch \eqref{e5} first.

We substitute the expansion
\begin{equation}
u = \sum_{i=0}^{\infty} u_i (y,z,t) \phi^{i+1}  , \label{e7}
\end{equation}
with $\phi$ given by \eqref{e3}, to the nonlinear equation \eqref{e2}, and collect terms with $\phi^{n-4}$, $n = 0, 1, 2, \dotsc$. In this way, we obtain recursion relations, which, for any given $n$, either determine an expression for $u_n (y,z,t)$ if $n$ is not a resonance, or lead to a compatibility condition if $n$ is a resonance. For $n = 0, 1, 2, 3$, which are resonances, we find that the compatibility conditions are satisfied identically and the functions $u_0 (y,z,t), u_1 (y,z,t), u_2 (y,z,t), u_3 (y,z,t)$ remain arbitrary. Then, for $n = 4, 5, 6, 7, 8, 9$, which are not resonances, the recursion relations give us explicit expressions for the coefficients $u_4, u_5, u_6, u_7, u_8, u_9$. The expressions are very complicated, not suitable for presentation in the paper, and one definitely needs to use computer algebra tools to obtain them. Finally, at the highest resonance $n = 10$, where  the function $u_{10} (y,z,t)$ remains arbitrary, a huge nontrivial compatibility condition appears, which contains 337 terms and involves the parameters $a$ and $b$, the functions $u_0 (y,z,t), u_1 (y,z,t), u_2 (y,z,t), u_3 (y,z,t)$ and their derivatives, and derivatives of the function $\psi (y,z,t)$.

The analysis of this compatibility condition at the highest resonance of the branch \eqref{e5} is hard but possible with the help of computer algebra tools. We find that the compatibility condition is satisfied identically, for arbitrary functions $u_0 (y,z,t), u_1 (y,z,t), u_2 (y,z,t), u_3 (y,z,t)$ and $\psi (y,z,t)$, in only three distinct cases of values of the parameters $a$ and $b$:
\begin{equation}
a = -10 , \qquad b = 0 , \label{e8}
\end{equation}
or
\begin{equation}
a = 5 , \qquad b = \pm 5 \sqrt{3} . \label{e9}
\end{equation}
For any other values of the parameters $a$ and $b$, the compatibility condition is not satisfied identically, the expansion \eqref{e7} must be modified by additional logarithmic terms (starting from the term proportional to $\phi^{11} \log \phi$), and this is a clear symptom of non-integrability.

To study the branch \eqref{e6}, we use the expansion
\begin{equation}
u = \sum_{i=0}^{\infty} u_i (y,z,t) \phi^{i+2} \label{e10}
\end{equation}
with $\phi$ given by \eqref{e3}, substitute it to the nonlinear equation \eqref{e2}, collect the terms with $\phi^{n-2}$, $n = 0, 1, 2, \dotsc$, and obtain the recursion relations which determine the coefficients $u_n (y,z,t)$ of the expansion (outside the resonances) and the compatibility conditions (at the resonances). We do not initially impose the already obtained conditions \eqref{e8} and \eqref{e9} on the parameters $a$ and $b$, because exactly the same conditions for $a$ and $b$ follow from the compatibility condition at the resonance 5 of this branch \eqref{e6}, whereas the corresponding expressions are sufficiently simple for presentation in the paper.

For $n = 0$ and $n = 1$ of the branch \eqref{e6}, we have resonances, the corresponding compatibility conditions are satisfied identically, and the functions $u_0 (y,z,t)$ and $u_1 (y,z,t)$ remain arbitrary. For $n = 2$, $n =3$ and $n = 4$, we obtain, respectively, the following expressions for the coefficients of the expansion \eqref{e10}:
\begin{gather}
u_2 = \frac{u_1^2}{2 u_0} - \frac{u_0}{60} \left( a \psi_y + b \psi_z \right) , \label{e11} \\
u_3 = \frac{u_1^3}{6 u_0^2} - \frac{u_1}{60} \left( a \psi_y + b \psi_z \right) \label{e12}
\end{gather}
and
\begin{gather}
u_4 = \frac{u_1^4}{24 u_0^3} + \frac{u_0 \psi_t}{40} - \frac{u_1^2}{120 u_0} \left( a \psi_y + b \psi_z \right) \notag \\
\qquad - \frac{u_1}{240 u_0} \left( a u_{0,y} + b u_{0,z} \right) + \frac{1}{240} \left( a u_{1,y} + b u_{1,z} \right) \notag \\
\qquad + \frac{u_0}{1440} \left( ( a - 10 ) \psi_y^2 - 2 b \psi_y \psi_z - ( a + 10 ) \psi_z^2 \right) . \label{e13}
\end{gather}
The case of $n = 5$ is a resonance, the function $u_5 (y,z,t)$ remains arbitrary, and the following nontrivial compatibility condition appears:
\begin{equation}
\left( a^2 + 5 a - 50 \right) \psi_{yy} + \left( 2 a b - 10 b \right) \psi_{yz} + \left( b^2 - 5 a - 50 \right) \psi_{zz} = 0 . \label{e14}
\end{equation}
This compatibility condition is satisfied identically, for any function $\psi (y,z,t)$, if and only if the parameters $a$ and $b$ satisfy the system of equations
\begin{equation}
( a + 10 ) ( a - 5 ) = 0 , \qquad ( a - 5 ) b = 0 , \qquad b^2 = 5 ( a + 10 ) , \label{e15}
\end{equation}
that is, exactly in the three distinct cases given by \eqref{e8} and \eqref{e9}.

It is nice that we have found an easier way to the conditions \eqref{e8} and \eqref{e9}, suitable for presentation in the paper. However, we still have to do a lot of computational work for the branch \eqref{e6}. For $n = 6, 7, 8, 9, 10, 11$, which are not resonances, the recursion relations give us expressions (very complicated, indeed) for the coefficients $u_6, u_7, u_8, u_9, u_{10}, u_{11}$ of the expansion \eqref{e10}. Then, at the highest resonance of this branch, that is for $n = 12$, where the function $u_{12} (y,z,t)$ remains arbitrary, we obtain a huge nontrivial compatibility condition, which contains 1861 terms. Finally, we verify that this compatibility condition is satisfied identically, for arbitrary functions $u_0 (y,z,t), u_1 (y,z,t), u_5 (y,z,t)$ and $\psi (y,z,t)$, whenever the parameters $a$ and $b$ satisfy \eqref{e8} or \eqref{e9}.

We can conclude now that the four-dimensional bilinear equation \eqref{e1} passes the Painlev\'{e} test for integrability in only three cases of its coefficients, when the parameters $a$ and $b$ are given by \eqref{e8} or \eqref{e9}. These integrable cases, however, are neither new nor four-dimensional, actually. In the case of \eqref{e8}, we get from \eqref{e1} the three-dimensional bilinear equation
\begin{equation}
\left( D_x^6 - 10 D_x^3 D_y - 20 D_y^2 + 36 D_x D_t \right) \tau \cdot \tau = 0 , \label{e16}
\end{equation}
which is the well-known BKP equation \cite{DKM81}. In the two cases with $a$ and $b$ given by \eqref{e9}, we can write the bilinear equation \eqref{e1} in the form
\begin{equation}
\left( D_x^6 + 5 D_x^3 ( D_y \pm \sqrt{3} D_z ) - 5 ( D_y \pm \sqrt{3} D_z )^2 + 36 D_x D_t \right) \tau \cdot \tau = 0 , \label{e17}
\end{equation}
where the $\pm$ signs correlate with the sign in \eqref{e9}. The linear transformation of two independent variables
\begin{equation}
y' = y \mp \sqrt{3} z , \qquad z' = z \mp \sqrt{3} y \label{e18}
\end{equation}
turns the (formally) four-dimensional equation \eqref{e17} into the three-dimensional BKP equation \eqref{e16}, with $y$ replaced by $y'$ and without any derivative with respect to $z'$ in it. In this sense, all the Painlev\'{e}-integrable cases of the four-dimensional bilinear equation \eqref{e1} are effectively three-dimensional and equivalent to the BKP equation.

\section{Extra equations} \label{s3}

The $D_4^{(1)}$-associated bilinear equations of the lowest degree 6 are linear combinations of the following three linearly independent equations \cite{KW89,CVY18}:
\begin{gather}
\left( D_x^6 + 36 D_x D_t - 10 D_y^2 - 10 D_z^2 \right) \tau \cdot \tau = 0 , \label{e19} \\
\left( D_x^3 D_y + D_y^2 - D_z^2 \right) \tau \cdot \tau = 0 , \label{e20} \\
\left( D_x^3 D_z - 2 D_y D_z \right) \tau \cdot \tau = 0 . \label{e21}
\end{gather}
(It is not said in \cite{KW89,CVY18} what are the coefficients of those linear combinations, and we guess that any.) In Section~\ref{s2}, we have studied the integrability of all the four-dimensional linear combinations \eqref{e1}. Now, for completeness, let us briefly consider the singularity analysis of the remaining linear combinations, initially three-dimensional ones. They are the bilinear equation
\begin{equation}
\left( D_x^3 D_y + D_y^2 - D_z^2 \right) \tau \cdot \tau + c \left( D_x^3 D_z - 2 D_y D_z \right) \tau \cdot \tau = 0 , \label{e22}
\end{equation}
with a parameter $c$, and the bilinear equation \eqref{e21} itself.

For the three-dimensional fourth-order equation \eqref{e22}, there is one branch to be studied. It describes the first-order zeroes of solutions near any non-characteristic hypersurface, the resonances being $-1, 0, 1, 6$. The compatibility conditions at the resonances 0 and 1 are satisfied identically. However, the nontrivial compatibility condition (with 100 terms) at the resonance 6 is satisfied identically for only two values of the parameter $c$,
\begin{equation}
c = \pm \frac{1}{\sqrt{3}} . \label{e23}
\end{equation}
For $c$ given by \eqref{e23}, we can write the bilinear equation \eqref{e22} in the form
\begin{equation}
\left( D_x^3 \Bigl( D_y \pm \frac{1}{\sqrt{3}} D_z \Bigr) + \left( D_y \mp \sqrt{3} D_z \right) \Bigl( D_y \pm \frac{1}{\sqrt{3}} D_z \Bigr) \right) \tau \cdot \tau = 0 , \label{e24}
\end{equation}
where the $\pm$ and $\mp$ signs correlate with the sign in \eqref{e23}. The linear transformation of two independent variables
\begin{equation}
y' = - \frac{1}{2} y \pm \frac{\sqrt{3}}{2} z , \qquad z' = z \pm \sqrt{3} y \label{e25}
\end{equation}
turns the bilinear equation \eqref{e24} into the bilinear equation \eqref{e21}, in which $y$ and $z$ are replaced by $y'$ and $z'$, respectively. Consequently, all the Painlev\'{e}-integrable cases of the bilinear equation \eqref{e22} are equivalent to the bilinear equation \eqref{e21}, which also possesses the Painlev\'{e} property therefore.

It only remains to note that the bilinear equation \eqref{e21} appeared for the first time in \cite{I80} (see Appendix B there).

\section{Discussion} \label{s4}

In this paper, we have studied the integrability of the four-dimensional sixth-order bilinear equation \eqref{e1} associated with the exceptional affine Lie algebra $D_4^{(1)}$. We have shown that this equation passes the Painlev\'{e} test for integrability in only three cases of its coefficients, given by \eqref{e8} and \eqref{e9}, exactly when the equation either coincides with the three-dimensional BKP equation \eqref{e16} or is related to the BKP equation by the linear transformation \eqref{e18} of two independent variables. Also, we have briefly considered the integrability of another $D_4^{(1)}$-associated bilinear equation, the initially three-dimensional equation \eqref{e22}, and transformed its integrable cases to the Ito equation \eqref{e21}.

The main result of this paper can be reformulated in the following way: every genuinely four-dimensional case of the bilinear equation \eqref{e1} fails to pass the Painlev\'{e} test for integrability. Therefore it could be an interesting problem to study multi-soliton solutions of the bilinear equation \eqref{e1} with any values of $a$ and $b$ different from \eqref{e8} and \eqref{e9}, in order to find any difference in the soliton dynamics between the genuinely four-dimensional cases and the three-dimensional integrable cases. Other integrability criteria, based on generalized symmetries, conservation laws, etc., could also be applied.

The last remark concerns the transformation \eqref{e18} which effectively merges two independent variables of the formally four-dimensional equation \eqref{e17} into one independent variable of the three-dimensional equation \eqref{e16}. Intentionally or not, the importance of such transformations is often ignored in the current literature. This generates a flood of ``novel'' higher-dimensional integrable equations which are nothing but the well-known old lower-dimensional ones with, say, $\partial_x$ replaced everywhere by $\partial_x + \partial_y$, or the like. Integrability of this kind could only be called ``true integrability in fake dimensions''.

\end{document}